\documentclass{optica-article}

\journal{opticajournal}
\articletype{Research Article}
\usepackage{comment}
\usepackage{lineno}

\begin{document}

\title{Observation of stability of Gaussian beams and off-axis beam-cleaning in graded-index media}

\author{Varun Sharma,\authormark{1,*} Nicholas Bender,\authormark{1}, Henry Haig,\authormark{1},Myungjoon Kim,\authormark{1},Demetrios N. Christodoulides,\authormark{2},and Frank W Wise\authormark{1}}

\address{\authormark{1}School of Applied and Engineering Physics, Cornell University, Ithaca, New York 14853, USA\\
\authormark{2}Ming Hsieh Department of Electrical and Computer Engineering, University of Southern California, Los Angeles, California 90089, USA}

\email{\authormark{*}vs527@cornell.edu} 


\begin{abstract*} 
While nonlinear effects in graded-index (GRIN) multimode fibers have been studied extensively, little is known about nonlinear effects in larger GRIN waveguides, where the number of modes approaches infinity and modal dispersion becomes negligible. Here we show that Gaussian beams remain nearly invariant even with large nonlinear phase accumulation and on- or off-axis trajectories in GRIN rods. In addition, spatially-complex beams can undergo self-cleaning to single-lobed profiles for both on- and off-axis trajectories. Numerical simulations exhibit the features observed in experiments, and a general interpretation of these results that makes connection to beam-cleaning phenomena observed in GRIN fibers is proposed. 

\end{abstract*}

\section{Introduction}

Graded-index (GRIN) optical media are essential building blocks of modern photonics, employed across imaging, beam-forming, and short-haul telecommunications applications \cite{agrawal2023GRIN_Media}, to name a few. The parabolic index profile in GRIN structures creates a quadratic spatial phase that leads to periodic self-imaging of the field, as would be produced by a series of lenses. The linear properties of GRIN waveguides underlie an attractive platform for studying nonlinear wave propagation: the parabolic index profile produces a unique spectrum of equally-spaced eigenmodes, and modal dispersion is small. These properties allow strong phase-matched nonlinear interactions with ultrashort light pulses. 

In the past decade, a variety of new nonlinear phenomena have been observed in GRIN multimode fibers, including control of spatiotemporal pulse evolution \cite{wright2015controllable,picozzi2015nonlinear}, direct observation of invariance of the self-imaging period with increasing intensity \cite{hansson2020nonlinear}, instabilities \cite{krupa2016observation_GPI,wright2016self_organized_instability}, beam self-cleaning \cite{krupa2017spatial_beam_cleaning}, and spatiotemporal mode-locking \cite{wright2017spatiotemporal_mode_locking}. New theoretical frameworks \cite{picozzi2014optical_thermodynamics,wu2019thermodynamic,podivilov2019hydrodynamic_2D_turbulance} have been introduced and help account for some of the observations.  While standard GRIN fibers with core diameters around 50~$\mu m$ have proven invaluable for studying nonlinear phenomena, the physics is still shaped by modal dispersion and interference. Quasi-periodic evolution of the field occurs, but true self-imaging does not. 
 
GRIN waveguide rods with millimeter-scale core diameters support on the order of $10^5$ transverse modes and have negligible modal dispersion. They exhibit nearly-perfect self-imaging, which allows them to guide Gaussian beams with off-axis trajectories. This contrasts with GRIN fibers, which only generate Gaussian output beams by pure excitation of the fundamental mode or with wavefront shaping. Nonlinear wave propagation in GRIN media has attracted quite a bit of theoretical attention \cite{bendow1981theory_CW_Beam_Propagation_Waveguide,manassah1988self_Foc_GRIN_Media,sammut1991gaussian,karlsson1992super_Gaussian_Approx,chen1992imaging_Gaussian_GRIN_rods,gagnon1991nonlinear,karlsson1992dynamics,yu1995spatio_temporal_soliton_GRIN,raghavan2000spatiotemporal,chien1996off_axis,longhi2003modulational,longhi2004self_focusing,agrawal2019invite_self_imaging,ahsan2020effect}. However, to our knowledge there are no experimental studies of nonlinear propagation in large GRIN waveguides. How nonlinear phenomena observed in GRIN fibers, such as beam cleaning, will translate to larger GRIN waveguides is unknown. Many nonlinear effects in GRIN fibers are fundamentally limited by modal dispersion and interference, and GRIN rods present a natural platform to explore these phenomena in the limit of vanishing modal dispersion. This regime may reveal generalizations of known effects, and it may host new nonlinear phenomena that are inaccessible in conventional GRIN fibers. 

Here we report a study of nonlinear pulse propagation in GRIN rods with about 1 mm core diameter – a factor of 10-20 larger than those of GRIN fibers. We find that Gaussian beams retain their shape and beam quality after propagation through GRIN rods even with both strong nonlinear phase accumulation and off-axis trajectories. We also observe self-cleaning of spatially-complex input beams with both on- and off-axis trajectories. In contrast to beam cleaning in multimode fibers, which results in power transfer to the fundamental mode (center) of the fiber, we find that 
beam cleaning in GRIN rods results in a single-lobed beam that maintains the trajectory (axial offset) determined by the launch condition.  Numerical simulations exhibit the main features of the experimental observations and provide valuable insight into the nonlinear propagation. We propose an interpretation of these phenomena in terms of the stability of single-lobed, near-Gaussian beams under nonlinear propagation in general GRIN media. 
\begin{figure}[ht]
     \centering
     \includegraphics[width = \columnwidth]{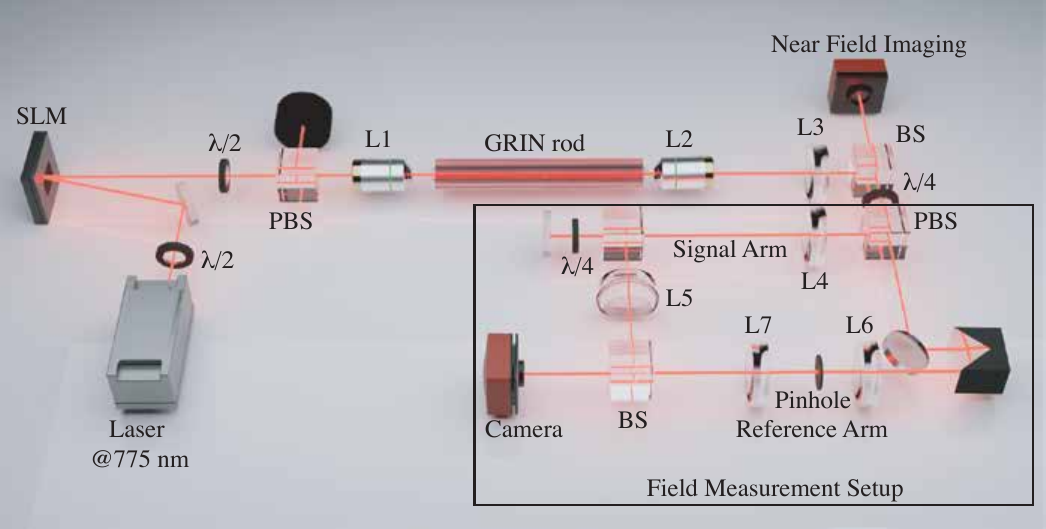}
     \caption{Schematic of the experimental setup, SLM: spatial light modulator, $\lambda$/2: half-wave plate, PBS: Polarizing beam splitter, L1-L2: Objective lenses, L2-L7: Lenses, $\lambda$/4: quarter-wave plate, BS: beam splitter}
     \label{fig:Fig 1}
\end{figure}

\section{Experiments and numerical simulations }
\subsection{Experimental setup}
The GRIN rods used in the experiment are standard commercial components (structures were obtained from Wavesource Photonics Inc.). They have a core diameter of 0.9~mm, a numerical aperture (NA) of 0.25, and are 15~cm long.  The self-imaging period determined experimentally is about 1~cm (Fig. 1 in Supplement 1). The structure supports around $10^5$ transverse spatial modes, so it is not effective (for theory or experiment) to work in the eigenmode basis. The exact nonlinear refractive index of the rod material is not known. From numerical simulations (discussed below) we roughly estimate that the nonlinear index is twice that of fused silica.

A schematic of the experimental setup is shown in Fig.~\ref{fig:Fig 1}. Pulses from a femtosecond fiber amplifier at 1550~nm are frequency-doubled to provide 500~fs pulses at 775~nm, with peak power as high as 160~kW. The beam becomes slightly elliptical in the process of second-harmonic generation. With this pulse duration, effects of chromatic dispersion on propagation through a 15~cm GRIN rod are negligible. The pulses diffract from a spatial light modulator (SLM) for wavefront shaping as desired. The SLM plane is imaged onto the back focal plane of an objective lens with numerical aperture of 0.15 to launch a phase-modulated beam onto the input face of the GRIN rod. Translation stages allow full control over the beam position and size at the input facet. The combination of a half-wave plate and a polarizing beam splitter (PBS) controls the input power. The field at the output of the GRIN rod is magnified 6$\times$ and imaged on a camera. To record the output electric field, a self-referenced Mach-Zehnder interferometer is implemented (Fig.~\ref{fig:Fig 1}), where one arm generates a plane wave reference beam using a Fourier filtering setup, and the other arm captures the signal through a 4f imaging system at the output face of the GRIN rod. The delay between the arms is controlled with 150~nm step size. This configuration provides the complete spatio-spectral characteristics of the output field \cite{pariente2016space_time_charac}. 
\begin{figure}[ht]
     \centering
     \includegraphics[width = 0.9\columnwidth]{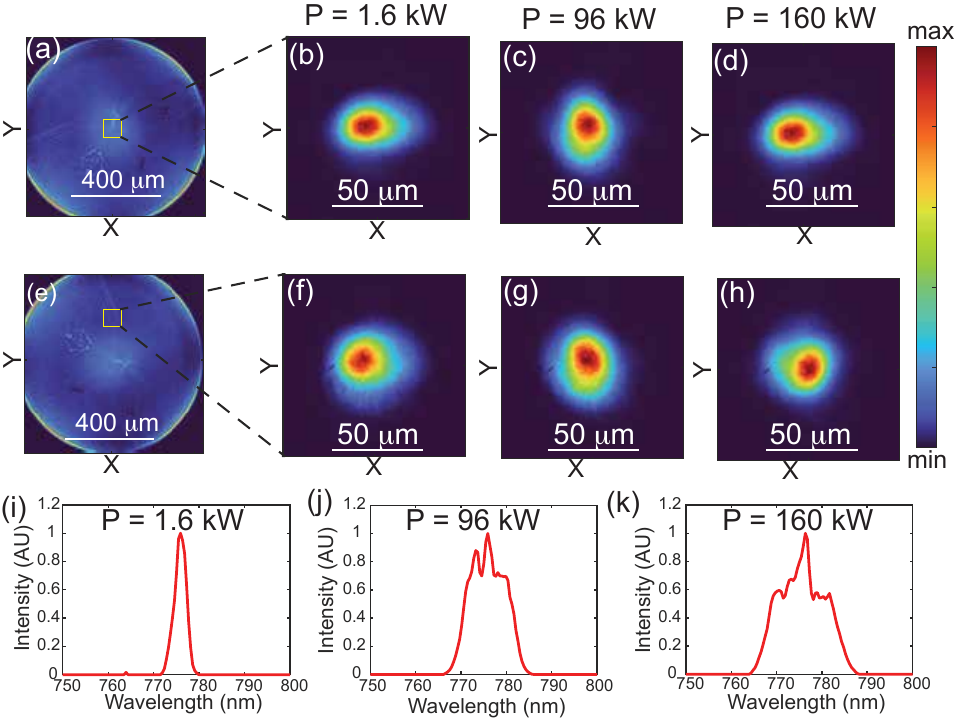}
     \caption{Gaussian beam propagation in GRIN rods. (a): GRIN rod output face with on-axis excitation. (b-d): Near-field intensity profiles at indicated input powers, (e): GRIN rod output face with off-axis excitation. (f-h): Near-field intensity profiles at indicated input powers. (i-k): output spectra with indicated powers.}
     \label{fig:Fig 2}
\end{figure}

\subsection{Experimental results}
As a first step, we investigated the nonlinear propagation of Gaussian beams in GRIN rods. Nonlinear propagation of on-axis \cite{karlsson1992super_Gaussian_Approx,karlsson1992dynamics,longhi2004self_focusing} and off-axis \cite{chien1996off_axis,ahsan2020effect} Gaussian beams have been analyzed theoretically. A beam with a full-width at half-maximum (FWHM) of 42~$\mu$m was launched on the axis of the GRIN rod. At low input power (1.6~kW), the beam undergoes self imaging evolution (Fig.~1(a) in Supplement 1) and the output beam retains the Gaussian profile (Fig.~\ref{fig:Fig 2}(b)), with an $M^2$ value of 1.13. With increasing power (Figs.~\ref{fig:Fig 2}(c,d)), the output profile remains nearly unchanged, with an $M^2$ of 1.15 at 160~kW. For off-axis excitation, the input beam is laterally shifted by $\sim$200~$\mu$m. As expected, the field follows a curved trajectory (Fig.~1(b) in Supplement 1). The output profile remains Gaussian at all input powers (Figs.~\ref{fig:Fig 2}(f,g,h)), with an $M^2$ value between 1.1 (1.6~kW) and 1.2 (160~kW). Thus, we conclude that Gaussian beams are stable in nonlinear propagation, as expected theoretically \cite{karlsson1992dynamics,karlsson1992super_Gaussian_Approx,ahsan2020effect}. At 1.6~kW, the output spectrum closely matches the input spectrum, with a FWHM width of 3~nm (Fig.~\ref{fig:Fig 2}i).  With increasing power significant spectral broadening occurs (Figs.~\ref{fig:Fig 2}(j,k)), corresponding to a nonlinear phase shift around 2$\pi$ at 160~kW. On-axis and off-axis input beams yield similar output spectra, which implies that nonlinear phase accumulation is independent of the launch conditions. This observation is consistent with the theoretical study \cite{ahsan2020effect}, which shows that the nonlinear enhancement factor remains the same for on-axis and off-axis Gaussian beams; however, a variation is predicted for different spatial beam shapes with the same effective area.

\begin{figure}[ht]
     \centering
     \includegraphics[width = \columnwidth]{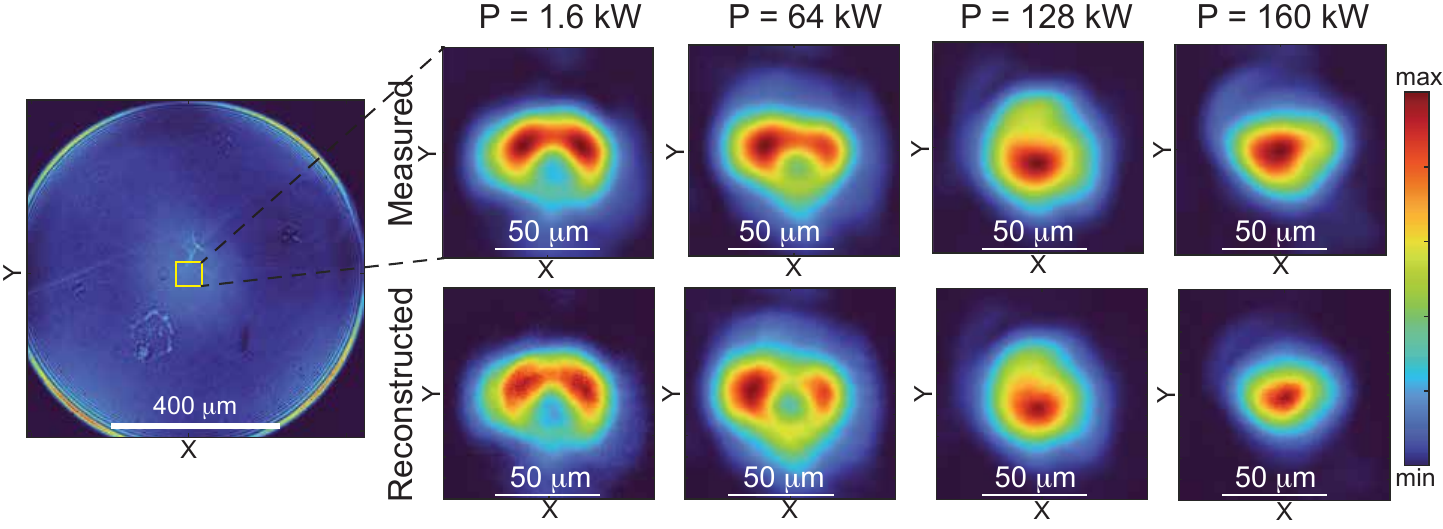}
     \caption{On-axis complex spatial beam evolution. Left: GRIN rod output face with location of the output beam. Right, top row: Near field camera measurements with varying output power. Right, bottom row: Corresponding intensity reconstructions via field measurement setup.}
     \label{fig:Fig 3}
 \end{figure}
To investigate the evolution of spatially-complex beams, 2x2 random phase patterns were encoded onto the SLM for experiments with on-axis and off-axis illumination. Representative results for on-axis excitation are summarized in Fig.~\ref{fig:Fig 3}. At low input power, the output beam profile is complex. At 64~kW the spatial distribution begins to change (Fig.~\ref{fig:Fig 3}d). By 128~kW a clear transition to a single lobe occurs (Figs.~\ref{fig:Fig 3}(f,g)), and this is maintained at 160~kW (Figs.~\ref{fig:Fig 3}(h,i). In all cases the reconstructed profile shows good agreement with the measured intensity, which validates the reconstructions. Field reconstruction allows us to look at the spatiospectral complexity of the output (Supplement 1). The strong breathing that accompanies beam cleaning (Figs.~2(a-d) in Supplement 1) in GRIN rods sits in stark contrast with the results of prior studies. 

\begin{figure}[h]
     \centering
     \includegraphics[width = \columnwidth]{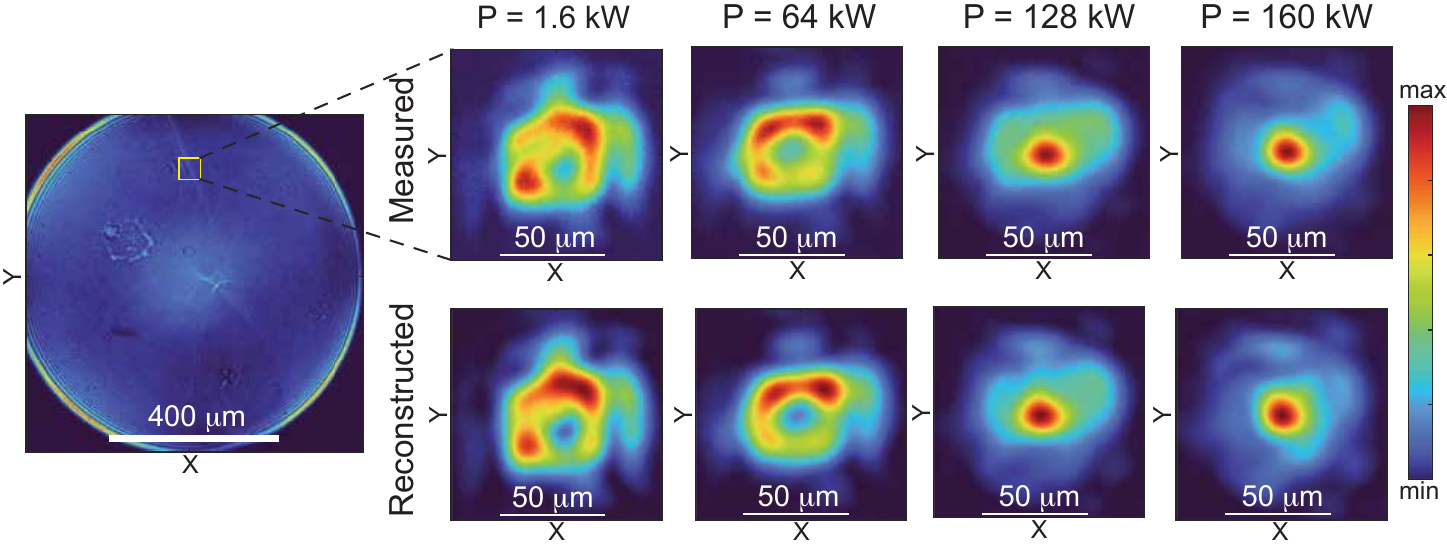}
     \caption{Off-axis complex spatial beam evolution. Left: GRIN rod output face with location of the output beam. Right, top row: Near field camera measurements with varying output power, Right, bottom row: Corresponding intensity reconstruction via field measurement setup. }
     \label{fig:Fig 4}
\end{figure}

Experimental studies of nonlinear propagation with off-axis launching of the input beam are quite limited. Hansson et al. imaged the "zig-zag" evolution of a Gaussian beam launched at a small angle with respect to the fiber axis \cite{hansson2020nonlinear}, and an experiment with off-axis complex beams showed that beam cleaning degrades with increasing offset from the center of the fiber \cite{krupa2019multimode_ST_Avenue}. In our experiments, phase-modulated beams were translated by $\sim$200~$\mu$m from the center of the rod (Fig.~\ref{fig:Fig 4}(a)) to study the off-axis evolution of complex input beams. The output beam profile measured at low power is complex (Fig.~\ref{fig:Fig 4}(b)). The behavior of the output beam with increasing power is very similar to the behavior when the field is launched on-axis. At 64~kW the beam just begins to change, while at 128~kW and above the beam exhibits an intense central lobe surrounded by a lower-intensity background (Fig.~\ref{fig:Fig 4}). Again, the reconstructed profiles agree with the measured intensities. It is important to keep in mind that the image panels in Fig.~\ref{fig:Fig 4} correspond to the off-axis location on the output face; the beam does not concentrate to the center of the GRIN rod, but rather to the off-axis position (Figs.~2(e-h) in Supplement 1). While the beam self-cleans, its centroid evidently follows the linear trajectory. The cleaned beams for on- (Fig.~\ref{fig:Fig 3}) and off-axis (Fig.~\ref{fig:Fig 4}) launching conditions should not be compared directly, because the launched beams did not have the same phase masks. These experimental results are in strong contrast to the prior study of off-axis excitation of GRIN fiber \cite{krupa2019multimode_ST_Avenue}, where significant cleaning was only observed with an on-axis beam.   
As is the case in beam cleaning in multimode fibers, there is a limit to the complexity of input beams that will undergo the self-cleaning shown above. With the current setup, experiments with randomly phase-modulated beams with 4x4 pixels reveal changes in the beam shape with increasing power, but not the transition to an intense single lobe (Supplement 1). It is important to note that the propagation distance studied here is quite limited by the available GRIN structures; at the highest input power, the propagation length is approximately 6 nonlinear lengths. What happens beyond this length is unkown.

The output beam quality may contribute to understanding the observed behavior, and will be of interest for potential applications. Measurements of the $M^2$ beam quality parameter \cite{siegman1998maybe_beam_quality, ross2013laser_beam_quality_metric} are shown in Supplement 1. For both on- and off-axis input beams, there is a small but measurable improvement in $M_{x}^2$, while $M_{y}^2$ is unchanged. The measured values range from 2.5 to 3 for the self-cleaned beams.

\subsection{Numerical simulations}
Ideally, experiments would be performed with varying propagation lengths, but it is not possible to do such measurements without disturbing the launch conditions. The photographs of the beam (Supplement 1) show the main qualitative features as it propagates through the GRIN rods. To further understand the evolution we turn to numerical simulations.  For simplicity and to focus on the spatial properties of the beam we solve the Gross-Pitaevski equation including diffraction, the parabolic index profile, and the Kerr nonlinearity, and neglecting dispersion:
\begin{equation}
  \frac{\partial E}{\partial z}=\frac{i}{2 k} \nabla_{\perp}^2 E - \frac{i k \Delta (x^2 + y^2)}{R^2} E + i k_0 n_2|E|^2 E
  \label{Equation:Eq1}
\end{equation}
where $\Delta$= $(n_{0}^2-n_{clad}^2) /2n_{0}^2$ is the refractive index contrast, where $n_{0} = n(x=0,y=0) $ denotes the refractive index at the GRIN rod center, for a core radius of $R = 0.45~\mu m$, and a wavelength of  $\lambda = 775~nm$, with $k_0= 2 \pi / \lambda$ and $k= n_{0} k_0$. The Kerr coefficient is set to $ n_2 = 6.4 \times 10^{-20}~m^2 W^{-1} $. Of course, this approach neglects variation of the beam across the temporal profile. The equation is solved using the split-step Fourier method on a 2048 x 2048 grid with absorptive boundary conditions. The propagation step size is $ 5~\mu m$, and the total propagation length is 15~cm. 

Results of launching a spatially-complex beam into the GRIN rod at an off-axis position are shown in Fig.~\ref{fig:Fig 5}. (The on-axis case is shown in Supplement 1.) In linear propagation, the beam follows the oscillatory trajectory and exhibits periodic self-imaging behavior while preserving its original spatial profile. With 160~kW input power, the oscillatory trajectory remains, and the beam shape evolves toward a single lobe superposed on that trajectory. The last cycle of the self-imaging trajectory is expanded in Fig.~\ref{fig:Fig 5}. The transition to a single lobe while retaining the oscillatory trajectory agrees with experimental observations, and we conclude that periodic self-imaging evolution occurs even as the beam undergoes cleaning. 

\begin{figure}[h]
     \centering
     \includegraphics[width = 0.8\columnwidth]{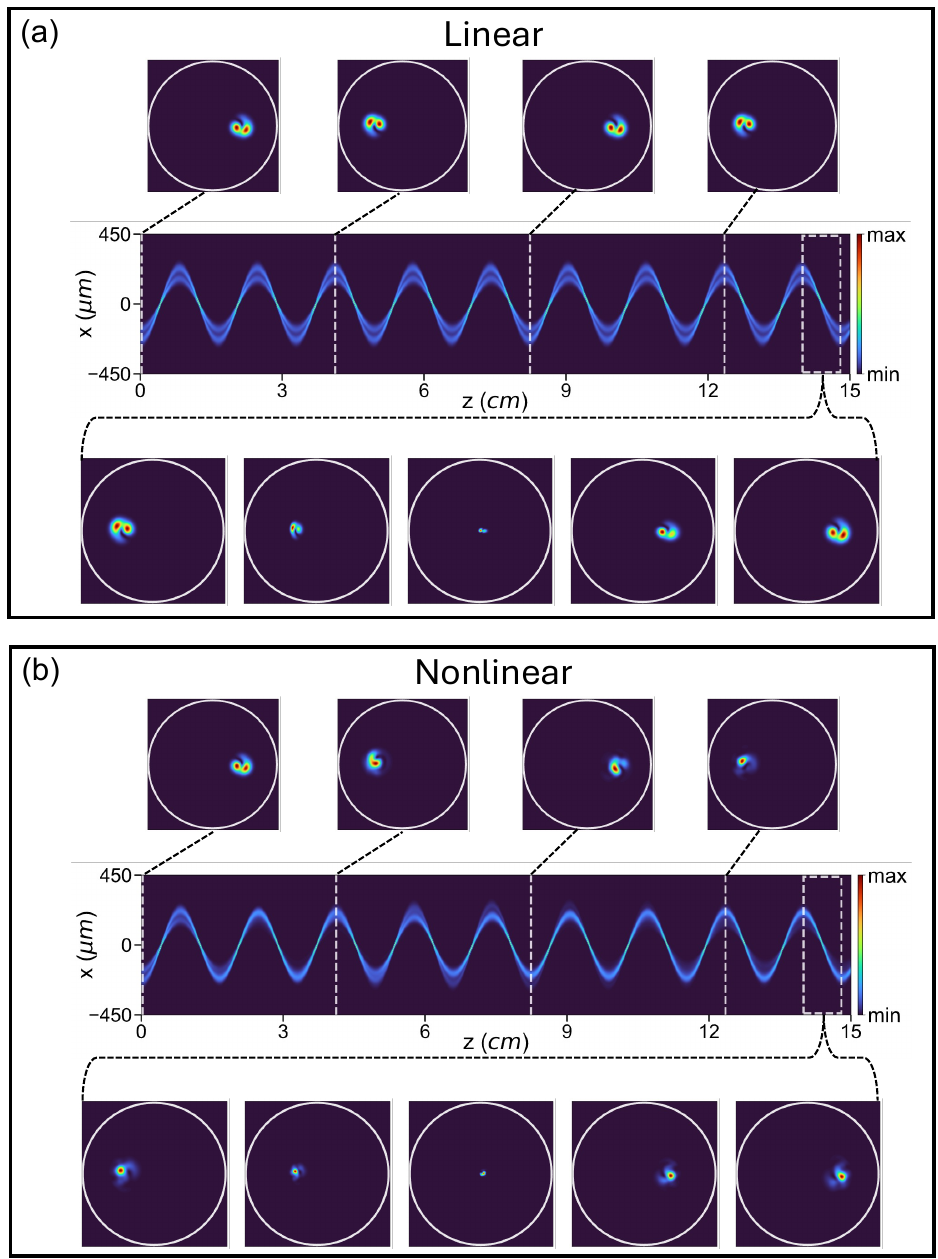}
     \caption{Numerical results for beam evolution in GRIN rod with spatially complex input beam launched at an off-axis position. (a): Cross-sectional view of beam intensity evolution (middle row) at 1~W input power, with spatial profiles at different propagation distances (top) and detailed evolution during the final self-imaging period (bottom). (b): Beam evolution at 160 kW input power. White circles represent the boundary of the core.
 }
     \label{fig:Fig 5}
\end{figure}

\section*{Discussion}

Although several phenomena reported here are not observed in GRIN fibers, the qualitative similarities between beam cleaning in GRIN rods and fibers suggest a common underlying mechanism. A variety of theoretical approaches have been applied to gain further understanding of beam cleaning. Numerical integrations of single-field \cite{krupa2017spatial_beam_cleaning} or multimode \cite{liu2016kerr_beam_cleaning} nonlinear Schrodinger equations exhibit the main features of prior beam-cleaning experiments. The first report of the effect was accompanied by a mean-field coupled-mode theory, which explained nonreciprocal power transfer from higher-order modes to the fundamental (i.e., most-intense) mode \cite{krupa2017spatial_beam_cleaning}.  A numerical study showed that spatial structure may decrease on propagation through GRIN fiber, accompanied by an increase of disorder in the temporal domain to conserve entropy \cite{laegsgaard2018spatial_beam_cleanup}. Classical wave condensation has been analyzed based on a discrete kinetic equation \cite{PhysRevA.83.033838,fusaro2019dramatic_acceleration_wave_condensation}, and beam cleaning was shown to be analogous to hydrodynamic turbulence in two dimensions \cite{podivilov2019hydrodynamic_2D_turbulance}. Optical thermodynamics shows that the equilibrium state of a nonlinear multimode wave is a superposition of transverse modes with a Rayleigh-Jeans distribution, with temperature determined by the input field \cite{ picozzi2014optical_thermodynamics,wu2019thermodynamic,ferraro2025negativeabsolutetemperatureattractor}. 

As far as we can tell,  none of the eigenmode-based theoretical frameworks account for the off-axis beam cleaning observed in GRIN rods. Even with on-axis excitation, the field undergoes strong breathing through periodic self-imaging (Supplement 1), which demonstrates that cleaning in GRIN rods does not necessarily transfer power toward a distribution dominated by the fundamental mode even under launch conditions that would do so in fibers. These discrepancies are not necessarily failures of the theories, which may not apply to the system studied here. For example, optical thermodynamics does not apply because the experimental conditions (number of modes, available power, and propagation distance) do not allow the field to equilibrate. It is possible that with further evolution, the field will approach the predictions of optical thermodynamics. 

We suspect that beam cleaning in GRIN rods is a version of the phenomenon observed in optical fibers generalized to the limit of an infinite number of modes and negligible modal dispersion. The work of Ahsan and Agrawal based on the use of a propagation kernel \cite{ahsan2020effect} neglects any influence of the temporal evolution on the self-imaging, which should be a good approximation for the experiments reported above. Thus, it seems to provide a framework for study of the nonlinear evolution of complex beam shapes in large GRIN structures, which would be interesting. Here, we offer a qualitative explanation of the experiments that seems to account for the main experimental facts. Our experiments show that Gaussian beams are stable in nonlinear GRIN media (Fig. \ref{fig:Fig 2}). This is theoretically-predicted for on- and off-axis excitation \cite{karlsson1992dynamics,karlsson1992super_Gaussian_Approx,ahsan2020effect}, and can be understood intuitively: the spatial nonlinear phase of a Gaussian beam has approximately the same profile as the linear phase that arises from the parabolic index. A Gaussian beam maintains its shape through a self-imaging period, so the accumulated nonlinear phase will also maintain its shape. Thus, a Gaussian intensity profile does not change. Beams with complex spatial profiles will accumulate correspondingly-complex nonlinear phase profiles, and will generally change shape as they propagate. If the spatial profile evolves into a Gaussian or single-lobed shape, it will maintain that shape from that point on. The evolution to a single spatial lobe will not happen with all input beams, and is less likely with greater complexity of the input beam. This qualitative framework allows us to reconcile the observations of cleaning in GRIN rods and fibers. Superpositions of higher-order modes can generate single lobes at specific axial positions, but they undergo strong evolution through an imaging period, which prevents stabilization. In fibers, only the fundamental mode maintains a single-lobed profile throughout the self-imaging period. Consequently, cleaning to a single-lobed spot that is off-axis is not observed.  

The results presented here suggest several avenues for future study. Studies of pulse propagation in longer GRIN rods and with higher power will naturally be interesting, to learn more about the asymptotic behavior. Systematic variation of the number of transverse modes through the core diameter could reveal a transition between fiber-like and rod-like cleaning behaviors. Development of a theoretical framework that encompasses both few-mode and many-mode limits would add valuable insight in this area, and would be useful for potential future applications of GRIN media, such as high-power pulse delivery.

\section*{Conclusion}
We have investigated the propagation of Gaussian and complex beams in GRIN rods with negligible modal dispersion. Under strong nonlinear conditions, Gaussian beams remain Gaussian regardless of the launch condition. Spatially-complex input beams, whether launched on- or off-axis, can undergo self-cleaning along their self-imaging trajectories. While qualitative arguments and numerical simulations are consistent with the experimental results, it will be valuable to have a systematic theoretical explanation that accounts for nonlinear propagation in all GRIN media.

\begin{backmatter}
\bmsection{Funding}
Department of the Navy, Office of Naval Research (N00014-20-1-2789), Simons Foundation. MK was supported by a fellowship from the Schmidt AI for Science program.

\bmsection{Acknowledgment} 
The authors acknowledge stimulating discussions with Yi-Hao Chen.

\bmsection{Disclosures} The authors declare no conflicts of interest.

\smallskip
\bmsection{Data Availability} Data underlying the results presented in this paper are not publicly available at this time but may be obtained from the authors upon request.

\bigskip

\bmsection{Supplemental document}
See Supplement 1 for supporting content.
\end{backmatter}

\bigskip

\bibliography{References}

\end{document}


\maketitle

\section{Beam propagation inside GRIN rods}
\begin{figure}[H]
     \centering
     \includegraphics[width = 0.6\columnwidth]{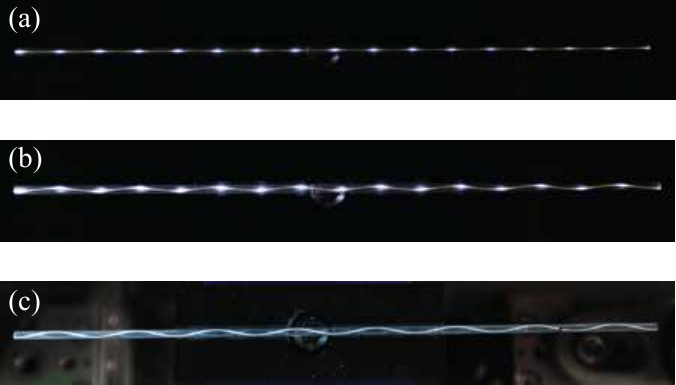}
     \caption{Ultrashort pulses inside GRIN rod (a): On-axis pulse propagation, (b,c): Off-axis pulse propagation}
     \label{fig:Fig S1}
 \end{figure}

At higher input powers, GRIN media exhibit multiphoton absorption \cite{hansson2020nonlinear} and generate visible fluorescence. Using a camera, we captured top-view images of the GRIN rods with excitation on-axis, off-axis, and close to the core boundary at 30~kW input power (Figs.~\ref{fig:Fig S1}(a,b,c) respectively). These images visually reveal the trajectory and self-imaging behavior of the pulses inside the GRIN rods.

\section{Beam cleaning and pulse propagation}

\begin{figure}[H]
     \centering
     \includegraphics[width = 0.8\columnwidth]{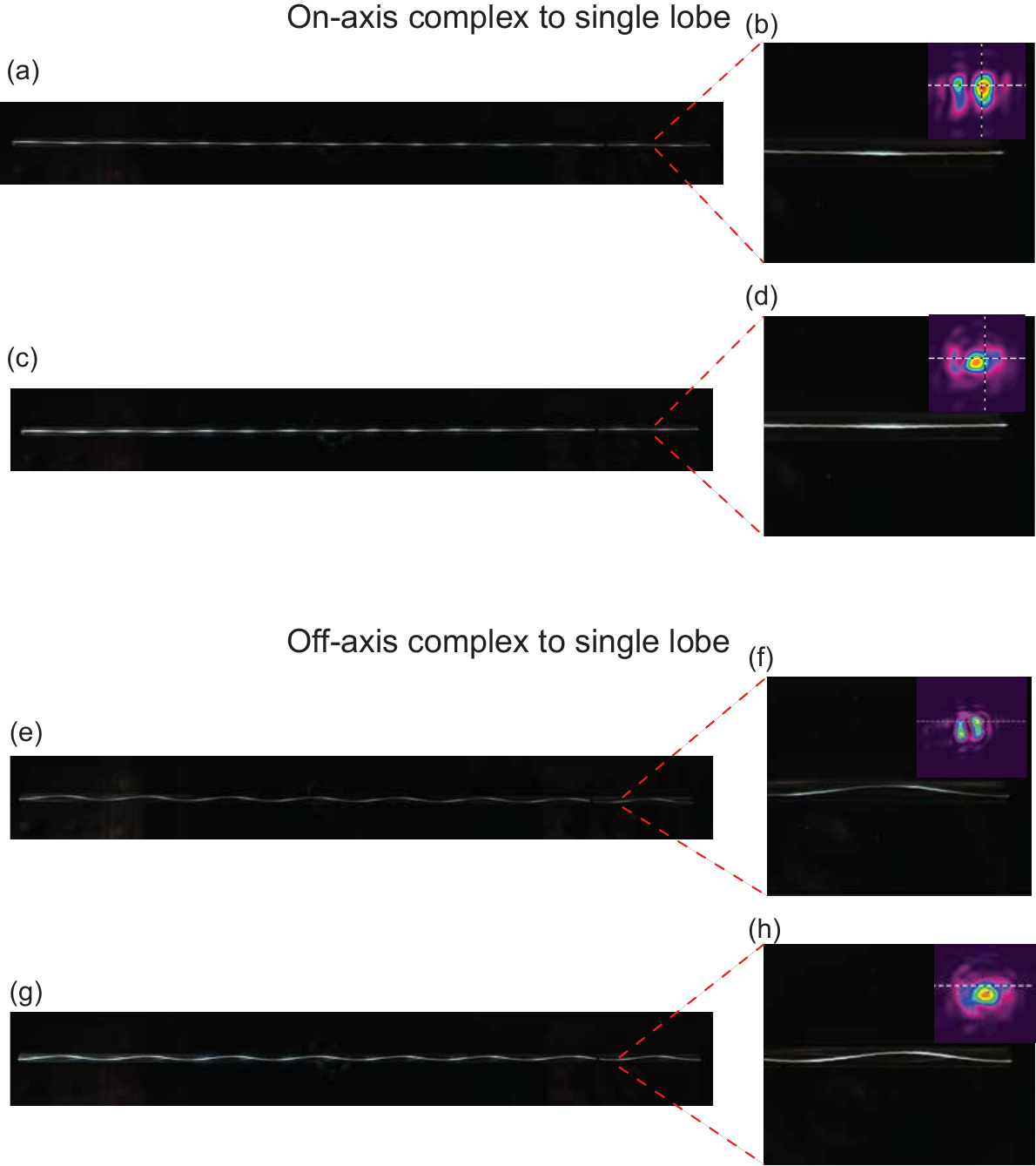}
     \caption{Beam cleaning and pulse propagation inside GRIN rod. (a-d): On-axis complex beam evolution to single lobe with beam trajectory. (e-h): Off-axis complex beam evolution to single lobe with beam trajectory.}
     \label{fig:Fig S2}
\end{figure}

Beam trajectories were imaged to capture the evolution from a complex input distribution to a cleaned beam at high power. For on-axis illumination, the beam path is shown in Fig.~\ref{fig:Fig S2}(a), with the final self-imaging period highlighted in the Fig.~\ref{fig:Fig S2}(b). The corresponding beam profile at 30 kW input power is shown in the inset of Fig.~\ref{fig:Fig S2}(b). As the power increases to 160 kW, the beam transitions to a single-lobed profile while preserving its on-axis breathing behavior, as illustrated in Figs.~\ref{fig:Fig S2}(c,d).

Off-axis illumination was also investigated. Figs.~\ref{fig:Fig S2}(e,f) present the beam trajectory and near-field intensity, including the last self-imaging period. At 160 kW, the beam similarly evolves into a single-lobed structure, while maintaining a curved trajectory as shown in Figs.~\ref{fig:Fig S2}(g,h).

\section{Spatial complexity and beam evolution}
To investigate how spatial complexity influences beam evolution, we controlled phase modulation using a spatial light modulator (SLM). The incident beam covered 500 pixels along both the X and Y axes of the SLM. We introduced spatial complexity by imposing an M×M grid of random phase patterns. For example, a 2×2 grid provides four degrees of freedom. As shown in Supplementary Figs.~\ref{fig:Fig S3}(a–c), such a configuration leads to beam evolution into a single spot as input power increases from 1.6~kW to 160~kW. We then increased the grid size to 4×4 to further enhance input complexity. Under these conditions, the beam still evolves nonlinearly with power, but fails to transform into a single spot, as shown in Supplementary Figs.~\ref{fig:Fig S3}(d–f). These results were confirmed through multiple measurements over different random patterns for each grid size. As the input complexity increases, we observe fewer instances of beam convergence to a single spot.
\begin{figure}[H]
     \centering
     \includegraphics[width = 0.6\columnwidth]{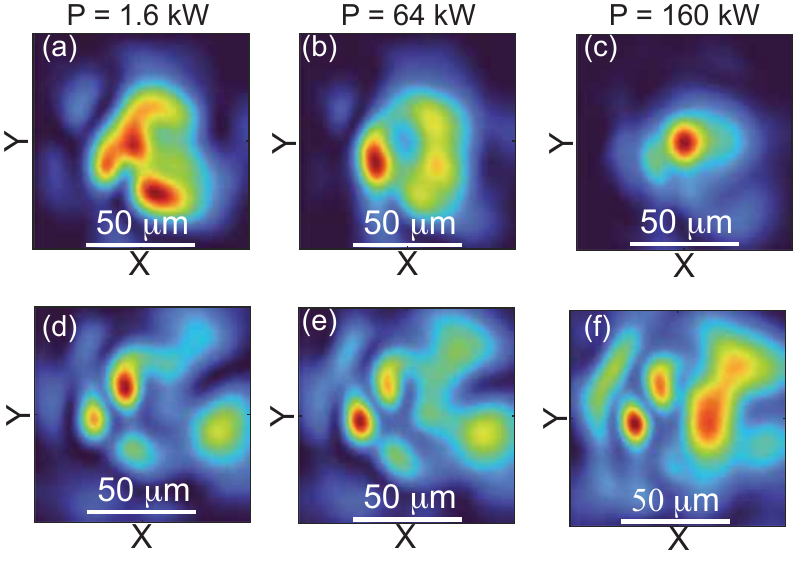}
     \caption{Output beam profiles with increase in input beam complexity, at indicated input powers. (a-c): Four degrees of freedom. (d-f): Sixteen degrees of freedom.}
     \label{fig:Fig S3}
 \end{figure}

\section{M$^2$ measurements with complex input beams}
\begin{figure}[H]
     \centering
     \includegraphics[width = 0.8\columnwidth]{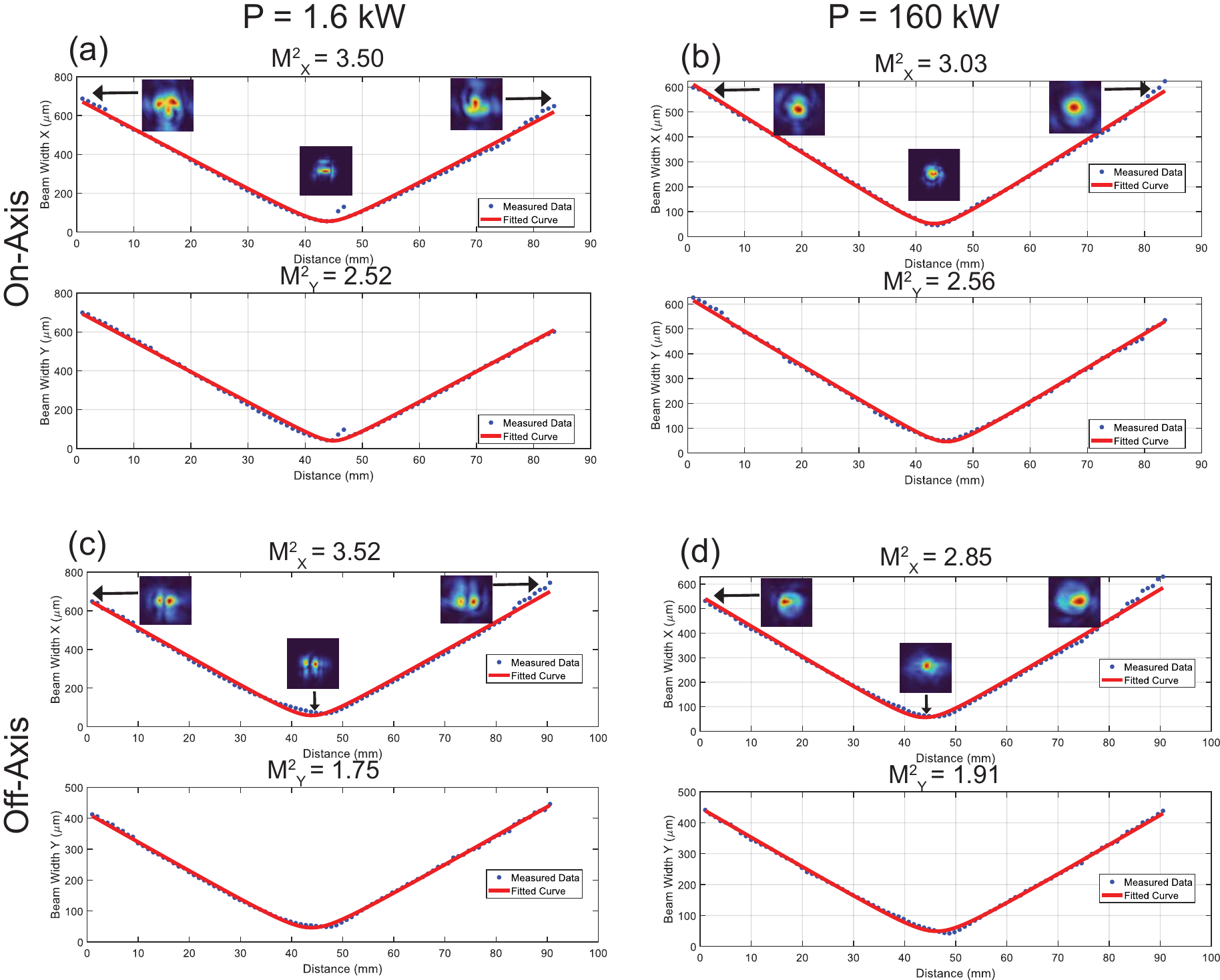}
     \caption{M$^2$ measurements with complex input beams.  (a,b): on-axis illumination. (c,d): off-axis illuminination. 
     Left column: M$^2$ measurement for 1.6~kW. Right column: M$^2$ measurement for 160~kW.}
     \label{fig:Fig S4}
 \end{figure}

Beam quality was assessed by performing $M^2$ measurements \cite{siegman1998maybe_beam_quality} for both on-axis and off-axis launch conditions. The output from the GRIN rod was collimated using a lens and then passed through a 200~mm focal length lens. Intensity profiles were recorded at 1~mm intervals across a 90~mm range near the focal region. The beam diameter was estimated using the $D4\sigma$ method, and the waist evolution was fitted to extract the $M^2$ values.
For the on-axis case, at 1.6~kW input power, $M^2$ values of 3.50 (X) and 2.52 (Y) were obtained. The corresponding intensity distributions near the focal plane and at both end points are shown in the insets of Fig.~\ref{fig:Fig S4}(a). At 160 kW, the beam forms a well-localized spot [Fig.~\ref{fig:Fig S4}(b)], with $M^2$ values of 3.03 (X) and 2.56 (Y).
In the off-axis case, measurements at 1.6~kW yielded $M^2$ values of 3.52 (X) and 1.75 (Y), with the corresponding intensity profiles shown in Fig.~\ref{fig:Fig S4}(c). At 160~kW input power, the beam again evolved into a single-lobed structure [Fig.~\ref{fig:Fig S4}(d)], with $M^2$ values of 2.85 (X) and 1.91 (Y).  In each case, the input beam has larger $M^2$ value for the x-axis than the y-axis and the value of $M^2$ for the x-axis improves a bit, while the value of $M^2$ value for the y-axis is constant.

\section{Spatio-spectral characteristics of output field}
\begin{figure}[H]
     \centering
     \includegraphics[width = 0.5\columnwidth]{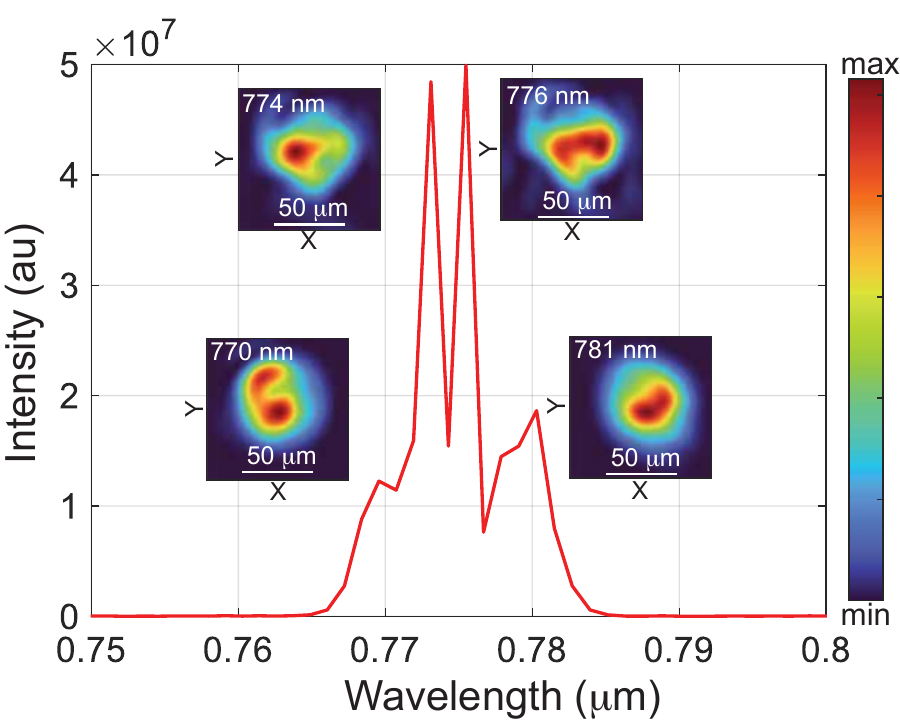}
     \caption{Example of spectral complexity of the output beam. Beam profiles are shown for indicated wavelengths, for the case of on-axis excitation at 160 kW power.}
     \label{fig:Fig S5}
\end{figure}

Measurements of the field with the self-referenced Mach-Zehnder interferometer reveal that the output field at high power is generally spatio-spectrally complex; i.e., the spatial profile varies across the spectrum. As an example, for an on-axis launched beam with 160 kW power (as in Fig. 3), the reconstructed spectrum is shown in Fig.~\ref{fig:Fig S5} along with the beam profiles recorded at the indicated wavelengths. 

\section{Numerical simulation for on-axis beam cleaning}
\begin{figure}[H]
     \centering
     \includegraphics[width = 0.8\columnwidth]{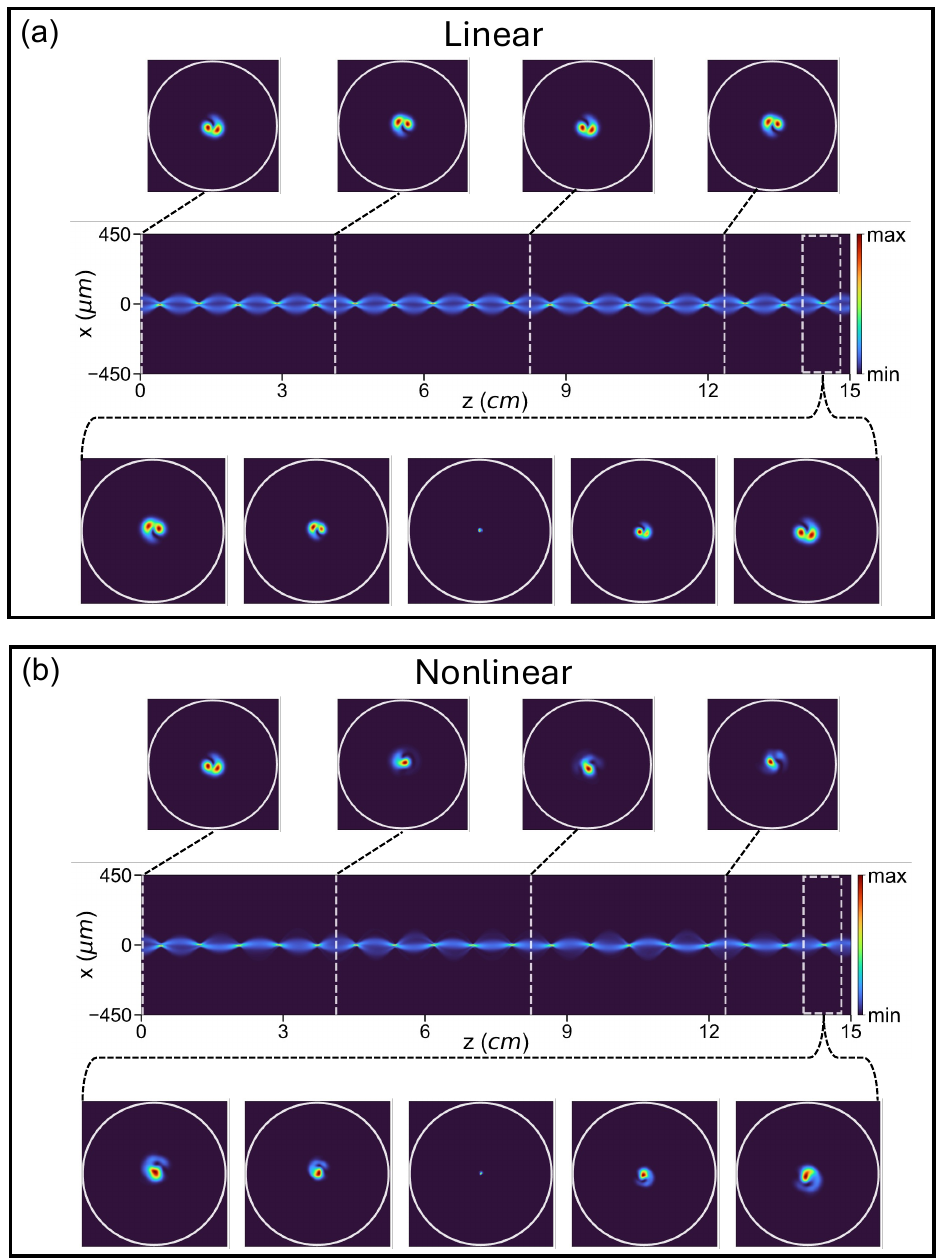}
     \caption{Numerical results for beam evolution in GRIN rod with spatially complex input beam launched on-axis. (a): Cross-sectional view of beam intensity evolution (center) at 1~W, with spatial profiles at different propagation distances (top) and detailed evolution during the final self-imaging period (bottom). (b): Same as (a) but for 160 kW input power. White circles represent the boundary of the core of the GRIN rod.}
     \label{fig:Fig S6}
\end{figure}

Fig.~\ref{fig:Fig S6} demonstrates the beam evolution for on-axis illumination of a GRIN rod in linear and nonlinear propagation regimes. With 160 kW input power, a multi-lobed intensity profile evolves to a single-lobed profile with strong periodic breathing in the beam size, as indicated in the expanded view of the last half-period of propagation.

\bibliography{References}